\documentclass[aps, prb, reprint, superscriptaddress, longbibliography, showkeys, citeautoscript, floatfix]{revtex4-2}

\usepackage[intlimits]{amsmath}
\usepackage{amssymb}
\usepackage{booktabs}
\usepackage{braket}
\usepackage{color}
\usepackage{float}
\usepackage[color=]{siunitx}
\usepackage{textgreek}
\usepackage{graphicx}
\usepackage{glossaries}
\usepackage{silence}
\usepackage[colorlinks=true,
            linkcolor=NavyBlue,
            citecolor=NavyBlue,
            filecolor=NavyBlue,
            urlcolor=NavyBlue
           ] {hyperref}
\usepackage[utf8]{inputenc}
\usepackage[version=4]{mhchem}
\usepackage{multirow}
\usepackage{soul} 
\usepackage{tabularx}
\usepackage[dvipsnames]{xcolor}
\usepackage[normalem]{ulem}
\usepackage{comment}
\usepackage{textcomp}

\usepackage{dcolumn}
\usepackage{float}
\usepackage{booktabs}
\usepackage{physics}

\setacronymstyle{long-short}
\newacronym{tmd}{TMD}{transition metal dichalcogenide}
\newacronym{pdms}{PDMS}{polydimethylsiloxane}
\newacronym{vdw}{vdW}{van der Waals}
\newacronym{dft}{DFT}{density functional theory}
\newacronym{vis}{VIS}{visible}
\newacronym{nir}{NIR}{near-infrared}
\newacronym{uv}{UV}{ultraviolet}
\newacronym{mse}{MSE}{mean squared error}
\newacronym{si}{SI}{Supporting Information}
\newacronym{wse2}{WSe\textsubscript{2}}{Tungsted diselenide}

\newcommand{\dtu}{
    Department of Electrical and Photonics Engineering, Technical University of Denmark,
    2800 Kgs. Lyngby,
    Denmark
}

\newcommand{\innsbruck}{
    Institut für Experimentalphysik, Universität Innsbruck,
    6020 Innsbruck,
    Austria
}

\newcommand{\jku}{
    Institute of Semiconductor and Solid State Physics, Johannes Kepler University Linz,
    4040 Linz,
    Austria
}

\begin{document}

\title{Exciton and biexciton preparation via coherent swing-up excitation in a GaAs quantum dot embedded in a micropillar cavity}

\author{Claudia Piccinini}
\altaffiliation{Contributed equally to this work}
\affiliation{\dtu}

\author{Aleksander Rodek}
\altaffiliation{Contributed equally to this work}
\affiliation{\dtu}

\author{Abdulmalik A. Madigawa}
\affiliation{\dtu}

\author{Ailton Garcia Jr.}
\affiliation{\jku}

\author{Saimon F. Covre da Silva}
\affiliation{\jku}

\author{Martin A. Jacobsen}
\affiliation{\dtu}

\author{Luca Vannucci}
\affiliation{\dtu}

\author{Gregor Weihs}
\affiliation{\innsbruck}

\author{Armando Rastelli}
\affiliation{\jku}

\author{Vikas Remesh}
\affiliation{\innsbruck}

\author{Niels Gregersen}
\email[]{ngre@dtu.dk}
\affiliation{\dtu}

\author{Battulga Munkhbat}
\email[]{bamunk@dtu.dk}
\affiliation{\dtu}

\begin{abstract}
Coherent control of quantum emitters is essential for scalable quantum photonic technologies. The recently proposed swing-up of quantum emitter (SUPER) scheme allows efficient and coherent preparation of single photons via off-resonant, red-detuned laser pulses, simplifying laser suppression and enhancing photon collection.
We present a systematic study of SUPER excitation applied to a single GaAs quantum dot in a low-Q micropillar cavity. We perform a comparison of the key figures of merit against the well-established two-photon excitation (TPE).
Despite requiring higher excitation powers, SUPER achieves near-unity population inversion of the exciton state ($\sim$95\%) and high single-photon purity ($g^{(2)}=0.03$) comparable to that under TPE, while also exhibiting a shortened decay time ($\sim$200 ps) reducing the time jitter in the exciton population.
A polarization-resolved analysis reveals that when both excitation and collection are aligned with one of the exciton dipoles, SUPER results in polarized single-photon emission, exceeding the resonant TPE saturation by a factor of 1.45.
Under optimized excitation conditions, we also observe biexciton preparation via a distinct SUPER resonance, confirmed by the appearance of the biexciton emission line, constituting the first experimental demonstration of biexciton preparation using SUPER. These findings are in good agreement with a proposed four-level theoretical model that incorporates the biexciton state. We also report that a slight misalignment of laser polarization induces an additional SUPER resonance that selectively populates the orthogonal exciton dipole, without altering the nominal excitation polarization. This unexpected behavior reveals a new degree of freedom for coherent state preparation. Our findings establish the SUPER scheme as a versatile tool for state-selective exciton and biexciton control.

\end{abstract}

\maketitle

\section{Introduction}
Highly efficient sources of pure and indistinguishable single photons are fundamental building blocks of photonic quantum information technologies~\cite{Heindel2023QuantumTechnology}. Among solid-state platforms, semiconductor quantum dots (QDs) are leading candidates for realizing such quantum light sources. However, the performance of QD-based single-photon sources is critically influenced by the choice of excitation scheme. Resonant excitation of the neutral exciton state is the established standard for generating highly indistinguishable single photons on demand~\cite{Somaschi2016, Ding2016, Nawrath2019a}. Despite its advantages, this technique requires complex polarization filtering or engineered photonic structures to suppress scattered laser light, which can limit photon extraction efficiency and scalability~\cite{Ates2009Post-SelectedMicrocavity, Wang2019, Uppu2020On-chipSource}.
To overcome these limitations, several off-resonant excitation protocols have been developed, including phonon-assisted excitation~\cite{Thomas2021BrightDipole}, two-photon excitation (TPE)~\cite{Muller2014On-demandPairs, Schweickert2018On-demandSource, Wei2022TailoringEmissions}, bichromatic excitation~\cite{HeNatPhys19, KoongPRL21, WilburAPLphot22, yan2025robustentangledphotongeneration}, and, more recently, the swing-up resonant excitation (SUPER) scheme~\cite{Bracht2021a, Vannucci2023HighlyExcitation, Karli2022SUPEREmitter}. These approaches enable coherent and efficient state preparation while reducing the need for stringent spectral filtering. 
The SUPER scheme, in particular, employs two red-detuned laser pulses to coherently “swing up” the QD from the ground state to an excited state. It has shown great potential for the direct exciton preparation on various QD platforms~\cite{Karli2022SUPEREmitter, Nawrath2019a, Boos2024CoherentSwingUp}. 
However, while SUPER has shown promise for exciton (X) preparation, a systematic experimental evaluation of its performance, particularly in comparison to other established excitation methods, is lacking. 
Important factors such as laser polarization alignment, spectral detuning, and the dynamics of exciton state preparation remain to be fully explored. 

Another key factor is the preparation of higher-order states, especially the biexciton (XX), which has been theoretically proposed~\cite{Bracht2023Dressed-stateSchemes} but has not been experimentally verified in the context of SUPER. This leaves open questions regarding their influence on exciton preparation dynamics and the potential of SUPER for generating entangled photon pairs via the XX–X cascade.

In this work, we present a comprehensive experimental investigation of the SUPER scheme applied to a single GaAs quantum dot embedded in a low-Q micropillar cavity. We first benchmark SUPER excitation against the well-known two-photon excitation, highlighting its advantages in terms of exciton preparation efficiency, single-photon purity, and temporal emission characteristics. Through polarization and time-resolved photoluminescence measurements, we identify optimal conditions for state-selective excitation, revealing that SUPER can coherently prepare either exciton dipole eigenstate with high polarization contrast and photon yield.
Crucially, we demonstrate the direct preparation of the biexciton state via SUPER, confirmed by unpolarized exciton emission resulting from the biexciton-exciton cascade. Finally, we report the emergence of an additional SUPER resonance condition that selectively populates the orthogonal exciton dipole without changing the excitation polarization, suggesting complex light–matter interactions beyond the standard three-level model and pointing to new directions for coherent quantum state control.


\section{Results}

\begin{figure*}
    \centering
    \includegraphics[width=0.85\textwidth]{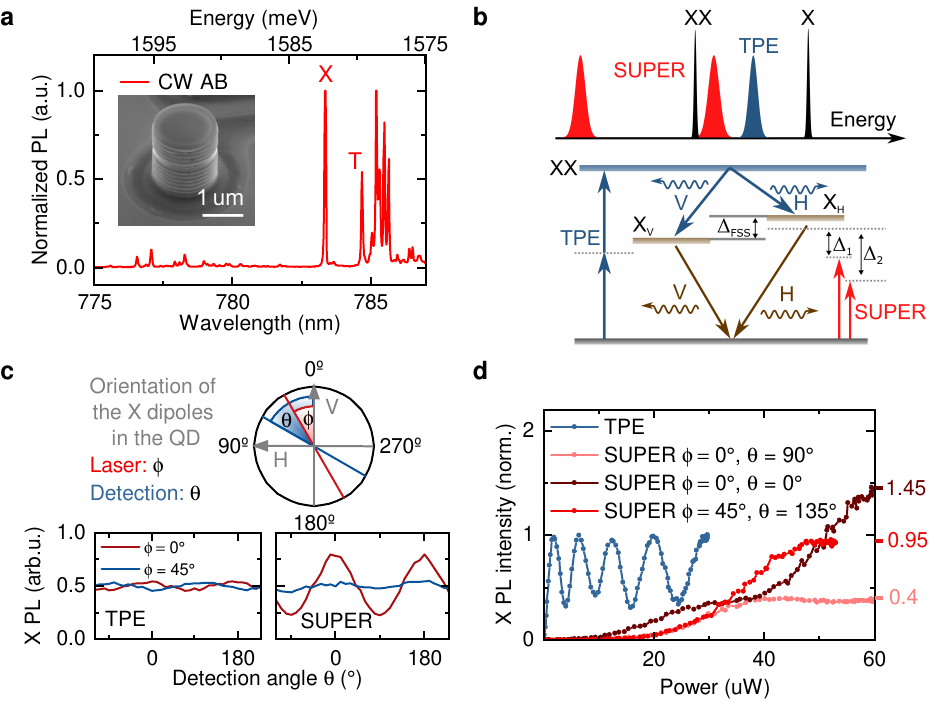}
    \caption{\textbf{a}~Photoluminescence (PL) spectrum of the examined QD under CW above-band excitation at 650 nm.
     Inset: Scanning electron microscopy (SEM) of the micropillar (D = 1.58 µm) hosting the QD. \textbf{b}~Illustration of the laser energies utilized for different excitation schemes in relation to the QD transitions and energy diagram schematics of the system. $\Delta_{\textrm{FSS}}\approx12$~$\mu$eV. Detuning from the neutral exciton energy: $\Delta_{\textrm{TPE}}$=-2 meV, SUPER pulses: $\Delta_1$=-3.5 meV, $\Delta_2$=-7.5 meV. \textbf{c}~Legend: Diagram of the employed convention for the excitation and detection angles in relation to the V,H-exciton states dipoles. Graph: Linear polarization anisotropy of the X PL under SUPER and TPE excitation for different orientations of the laser polarization. \textbf{d}~Comparison of the X PL intensity under TPE and SUPER for different polarization configurations. Data was normalized by the maximum number of counts under TPE excitation for linear polarization of detection.}
    \label{fig:Fig1}
\end{figure*}

The investigated device consists of the GaAs/AlGaAs quantum dot (QD) embedded in a low-Q micropillar cavity (Fig.~\ref{fig:Fig1}a). It was deterministically fabricated following the process described elsewhere~\cite{Madigawa2023AssessingDevices}, and further information on the micropillar cavity is provided in Supplementary Fig.~S1. All experiments were conducted at T=4~K in a closed-cycle cryostat equipped with nanopositioners and a low temperature objective lens (NA=0.82, 60$\times$). To generate laser pulses with tunable energy and linewidth for the excitation, we employed a folded-4f pulse shaper, incorporating a programmable spatial light modulator, similar to configuration described previously~\cite{Karli2022SUPEREmitter}. Further details on the experimental setup are provided in \textbf{Methods} section and Supplementary Fig.~S2.


\subsection{Polarization properties of SUPER-excited neutral exciton.}
The photoluminescence (PL) spectrum of the QD under above-bandgap excitation ($\lambda=650$ nm) is shown in Fig.~\ref{fig:Fig1}a. The dominant emission peak at $\lambda$ = 783.38 nm corresponds to the neutral exciton (X), while a weaker emission line at $\lambda$ = 785.38 nm is attributed to the biexciton (XX), with a corresponding binding energy of $E_b\sim$4.0 meV. Additional emission lines at longer wavelengths originate from multiexciton complexes, whereas weaker features at higher energies correspond to p- and d-shell transitions~\cite{Reindl2019HighlyDots}. In this work, we focus on the neutral exciton state.

We begin by identifying the optimal conditions for swing-up (SUPER) excitation and compare its performance with the well-established two-photon excitation (TPE). In the TPE case, the energy of the excitation pulse is set to the average energy of the X and XX transition line with its wavelength $\lambda\approx784.38$~nm. In contrast, the SUPER scheme employs two red-detuned pulses, with detunings of  $\Delta_1=-3.5$~meV and $\Delta_2=-7.5$~meV relative to the X transition. The full width at half maximum (FWHM) of the excitation laser was fixed at 0.48~meV.
An illustration of the pulse energies relative to the QD level diagram is shown in the Fig.\ref{fig:Fig1}b, and representative PL spectra under both TPE and SUPER are provided in Supplementary Fig.~S3. We emphasize that both excitation pulses are required for generating the exciton emission; no emission is observed when only one pulse is present.
To confirm the single-photon nature of the emission from the investigated sample, we performed second-order autocorrelation measurements (more details in \textbf{Methods} section and Supplementary Fig.~S4). Both excitation schemes yield high single photon purity, with low $g^{(2)}(0) = 0.009 \pm 0.002$ for TPE and $g^{(2)}(0) = 0.03 \pm 0.01$ for SUPER. Furthermore, time-resolved PL measurements confirm a fast, monoexponential decay of the exciton signal under SUPER excitation, in contrast to the delayed emission signature of the biexciton-exciton (XX-X) cascade observed under TPE (Supplementary Fig.~S7).  

Having established direct exciton preparation and single-photon emission under SUPER excitation, we now investigate its polarization properties. 
A key advantage of the SUPER protocol is that it does not require stringent polarization filtering. In standard resonant excitation, cross-polarization filtering is typically used to suppress laser scattering, which inevitably restricts the detected emission to $\sim$50~$\%$~\cite{Ollivier2020ReproducibilitySources} for rotationally symmetric vertically emitting sources. Two-photon excitation (TPE) avoids this limitation by exciting the biexciton state with a linearly polarized pulse and detecting the emitted photon cascade. However, due to the spin selection rules in the XX–X cascade, the resulting emission is equally distributed between orthogonal polarizations, limiting the degree of polarization control. This process is illustrated in the energy level diagram in the Fig.~\ref{fig:Fig1}b.
In contrast, under SUPER scheme the utilized laser pulses can be polarized along the exciton dipole axis, allowing full excitation and detection of a single polarization component, thereby enabling $\sim$100~$\%$ polarized emission without polarization filtering. To verify this we present in Fig.~\ref{fig:Fig1}c polarization anisotropy measurements of the neutral exciton (X) photoluminescence (PL) under both TPE and SUPER excitation. The polarization of the excitation laser was oriented either along the vertical axis, coinciding with the orientation of one of the X transition dipoles ($\phi=0^\circ$), or at 45$^\circ$ with respect to it. As expected, TPE exhibits no polarization dependence, as the exciton emission results from a biexciton–exciton cascade that populates the H and V states with equal probability. However, under SUPER excitation, when the laser polarization is aligned with the V-dipole of the exciton, we observe a strong polarization contrast in the PL, confirming selective excitation of a single dipole. At $\phi = 45^\circ$ the PL becomes unpolarized, similar to the TPE, due to the simultaneous excitation of both H and V states.  

To now determine the efficiency of exciton preparation for different polarization configurations we perform saturation measurements comparing the observed brightness with respect to the TPE. These results are presented in Fig.~\ref{fig:Fig1}d. Under two-photon excitation, the X PL intensity exhibits Rabi oscillations, with peak count rates corresponding to full population inversion of the system. Here, the signal was detected after passing through a linear polarizer, and therefore the plot is normalized to half of the total unpolarized TPE emission. We note that this should also be the case for SUPER excitation of both H and V dipole states with $\phi=45^\circ$. For optimized excitation conditions ($\Delta_1 = -3.8$ meV, $\Delta_2 = -9$ meV, with a power ratio $P_2/P_1 \approx 1.3$) we observe a clear saturation of the exciton PL signal reaching brightness levels comparable to those achieved with $(n+1)\pi$-pulses in TPE. This result aligns with previous demonstrations of near-unity population inversion of the exciton state through the SUPER method~\cite{Karli2022SUPEREmitter}. Based on this finding, we now align both excitation and detection polarizations along the vertical axis ($\phi \approx 0^\circ$, $\theta = 0^\circ$). To optimize the emission under this orientation, we fine-tune the second pulse detuning to $\Delta_2 = -7.4$ meV. In the co-polarized configuration, SUPER excitation yields a higher count rate at saturation compared to both TPE and the previous configuration of SUPER with $\phi \approx 45^\circ$ and $\theta = 135^\circ$. However, it does not produce the expected doubling of the count rate, instead reaching an enhancement factor of approximately $\sim$1.45. 
Another unexpected feature is that, in the orthogonal detection configuration ($\phi \approx 0^\circ$, $\theta = 90^\circ$), we still observe substantial H-polarized emission, despite the excitation being purely V-polarized. Polarization-resolved PL measurements performed across a full range of excitation and detection angles further confirm this behavior (Fig.~S11). While the strongest polarization contrast is observed when the excitation is aligned along the H or V dipoles, residual emission in the orthogonal polarization persists across all $\phi$ angles.

These counter-intuitive behaviors could arise from several sources: (i) residual ellipticity of the excitation laser, leading to unintended excitation of both dipoles; (ii) polarization mixing caused by finite fine-structure splitting (FSS); or (iii) more complex mechanisms involving higher-order states, such as biexciton (XX), that may lead to indirect population of the orthogonal dipole.
To rule out the first scenario, we quantified the ellipticity of the excitation laser by measuring its back-reflected signal from the planar region next to the micropillar. We confirm a high polarization extinction ratio approaching $10^2$ (Fig.~S12), strongly suppressing any significant excitation of the orthogonal dipole. 
To examine the second hypothesis, we modeled the system considering the fine structure splitting and the exciton linewidth of the QD. As shown in Fig.~S11, although the model qualitatively reproduces the experimental anisotropy maps, it predicts no orthogonal emission for any value of FSS when the excitation is aligned with one of the exciton dipoles. This suggests that polarization mixing from FSS alone cannot account for the observed orthogonal emission, which leaves the involvement of higher-order states in the excitation process.

\subsection{SUPER excitation within a four-level system}
\begin{figure}[t]
    \centering
    \includegraphics[width=\columnwidth]{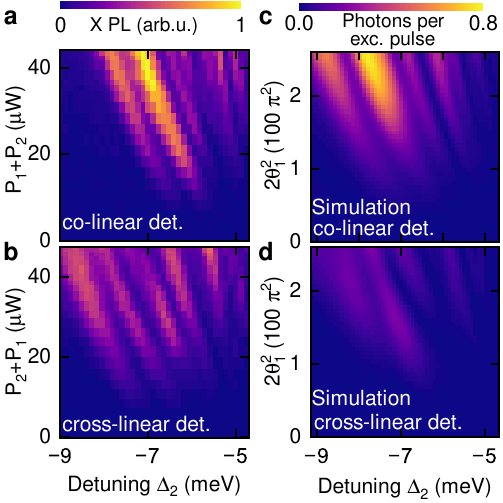}
    \caption{Experimental (\textbf{a,b}) and simulated (\textbf{c,d}) excitation power scans of the X emission intensity under SUPER scheme for different detunings $\Delta_2$ for co-(\textbf{a,c}) and cross-(\textbf{b,d}) linear configuration of detection polarization, $\phi\approx 0$. Excitation parameters: $\Delta_1=-3.7$~meV, P$_1$=P$_2$. Simulation parameters: binding energy of biexciton E$_{b}=4$ meV, linewidth of the pulses FWHM=0.48~meV, P$_1$=P$_2$, $\Delta_1=-3.7$~meV. 
}
    \label{fig:Fig2}
\end{figure}

To determine the origin of the observed polarization properties and their possible dependence on the particular excitation conditions we performed two-dimensional scans of the exciton PL intensity as a function of total excitation power and the detuning of the second pulse $\Delta_2$, with laser polarization oriented along the V-dipole and detection in either co- or cross-polarized configurations (Figs.~\ref{fig:Fig2}a–b). In the co-polarized case (Fig.~\ref{fig:Fig2}a), we see multiple resonance fringes, with the strongest emission appearing at $\Delta_2 \approx -7.4$ meV. 

The key experimental features—including both the overall intensity and the resonance fringes—are qualitatively well reproduced by the simulation shown in Fig.~\ref{fig:Fig2}c performed within a four-level-model that includes the ground state $\ket{G}$, two orthogonal exciton states $\ket{X_V}$ and $\ket{X_H}$ with linear polarization, and the biexciton state $\ket{XX}$. Their respective energies are $E_G = 0$, $E_{X_V} = \hbar \qty(\omega_0 - \frac 1 2 \Delta_{\rm FSS})$, $E_{X_H} = \hbar \qty(\omega_0 + \frac 1 2 \Delta_{\rm FSS})$, and $E_{XX} = \qty(2 \hbar \omega_0 - E_b)$, where $\hbar \Delta_{\rm FSS}$ is the fine structure splitting (FSS), and $E_b$ is the biexciton binding energy (more details in \textbf{Theory}). 

In the cross-polarized detection configuration, i.e., when the excitation is aligned with one of the exciton dipoles (see Fig.~\ref{fig:Fig2}b), where no emission is expected, we detect pronounced resonance fringes across the detunings. This result cannot be explained by a simplified three-level model consisting of only the ground and H/V exciton states (see details in Suppl. Fig.~S13b). In contrast, the four-level model that incorporates the biexciton (XX) state successfully reproduces the experimentally observed features, as shown in Fig.~\ref{fig:Fig2}d. This strongly indicates that biexciton population plays a critical role in enabling emission from the orthogonal exciton dipole. It is now clear, that under optimal excitation conditions, SUPER scheme can coherently drive both XX and X states, and this coherent interplay gives rise to the observed polarization mixing in the PL signal, which is highly sensitive to the excitation conditions. Importantly, these findings suggest a promising new route for coherent biexciton preparation using the SUPER protocol, representing, to the best of our knowledge, one of the first experimental demonstrations of this mechanism. In the next section, we provide direct experimental verification of this interpretation and explore the excitation conditions enabling efficient and selective preparation of the biexciton-exciton cascade under the SUPER scheme.

\subsection{Exciton Emission via Biexciton Cascade under SUPER}

\begin{figure*}
    \centering
    \includegraphics[width=0.85\textwidth]{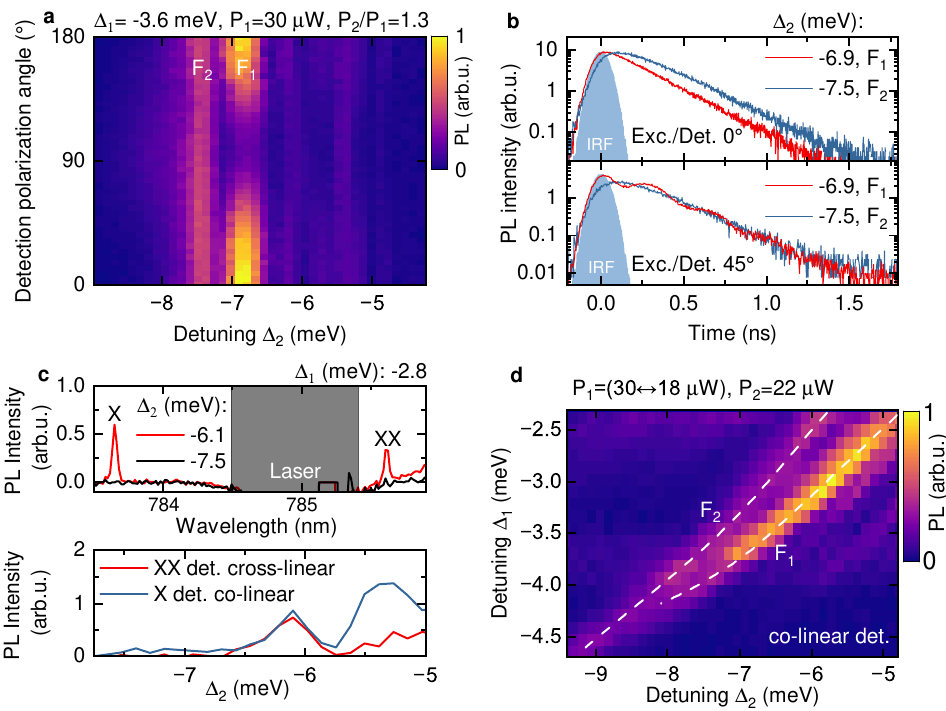}
    \caption{\textbf{a} Intensity of the neutral exciton (X) photoluminescence (PL) under SUPER excitation as a function of the detuning of the second excitation pulse and the detection angle of linear polarization. \textbf{b} Temporal decay of the X PL under SUPER excitation for different detunings of the second laser pulse, with the polarization of the laser aligned either along (upper panel) or between (lower panel) the dipole orientation of the exciton fine-structure-split states. \textbf{c} PL spectrum of the QD under SUPER excitation of the biexciton state (upper panel) and dependence of the X, XX intensity as a function of the detuning of the second pulse. (Data was normalized by the intensity observed at $\Delta_2=-6.1$ meV). \textbf{d} Intensity of the X PL under SUPER excitation while changing the detuning of the laser pulses. Dashed lines indicate the resonant conditions for exciting either the neutral- or the biexciton state. (P$_1$ is linearly increasing in the investigated range when tuning towards higher energies due to the limitation of the experimental setup.)}
    \label{fig:Fig3}
\end{figure*}

To experimentally verify and better understand the role of biexciton-mediated neutral exciton emission under SUPER excitation, we now investigate the emission polarization properties of the X  state as a function of the detuning of the second pulse. Figure~\ref{fig:Fig3}a shows the measured polarization anisotropy, obtained by scanning the detection polarization angle from 0° to 180°, while keeping the excitation laser fixed along the V-direction~(0$^\circ$). A clear difference emerges at two distinct detunings. At $\Delta_2 = -6.9$ meV (F$_1$), the PL is strongly polarized along the laser axis, consistent with selective excitation of the neutral exciton. In contrast, at $\Delta_2 = -7.5$ meV (F$_2$), the PL becomes unpolarized, resembling the behavior of biexciton–exciton cascade emission, as typically observed under TPE excitation.
This suggests that the two resonance fringes observed in Fig.~\ref{fig:Fig3}a correspond to SUPER-mediated excitation of the neutral exciton, either through direct population or indirectly via biexciton cascade. It also explains the excitation-power-dependent changes in polarization ratio observed in Fig.~\ref{fig:Fig1}d: as the laser power increases, the resonance fringes shift toward larger detunings (Fig.~\ref{fig:Fig2}), causing the SUPER process to switch between predominantly driving the exciton or the biexciton, depending on the fixed values of $\Delta_{1,2}$.

Further evidence linking the resonance at $\Delta_2~=~-7.5$~meV to the SUPER-mediated preparation of the biexciton is provided by the temporal dynamics of the neutral exciton emission. The upper panel of Fig.~\ref{fig:Fig3}b shows time-resolved PL traces for the two detunings corresponding to the resonances identified in Fig.~\ref{fig:Fig3}a. At $\Delta_2 = -7.5$~meV, where the emission is unpolarized, the PL decay exhibits a pronounced rise time and delay, consistent with biexciton-mediated population transfer, similar to what is observed under TPE excitation (Fig.~S7). In contrast, the resonance at $\Delta_2 = -6.9$ meV shows fast decay, characteristic of direct exciton excitation. Furthermore, in the lower panel of Fig.~\ref{fig:Fig3}b, we show the PL dynamics when the laser polarization is oriented between the H and V dipoles. Quantum beats~\cite{FlissikowskiPRL01} are observed for $\Delta_2 = -6.9$~meV but not for $\Delta_2 = -7.5$~meV , ruling out coherent superposition of the H and V exciton states for the latter condition and further supporting the interpretation of a biexciton-driven cascade.

To provide direct evidence of biexciton involvement under SUPER excitation, we aimed to directly detect photoluminescence from the biexciton (XX) state. To enable this, we adjusted the detuning of the first pulse from $\Delta_1 = -3.6$~meV to $\Delta_1 = -2.8$~meV, thereby avoiding spectral overlap between the excitation laser and the XX emission line, while still not within the range of phonon-assisted TPE. The upper panel of Fig.~\ref{fig:Fig3}c shows the resulting PL spectra for two different values of the second-pulse detuning. At $\Delta_2 = -6.1$~meV, both exciton (X) and biexciton (XX) emission are clearly observed. In contrast, at $\Delta_2 = -7.5$~meV, no PL from either state is detected, confirming that both excitation pulses are required for the process, ruling out resonant or phonon-assisted TPE via the spectral tails of the pulses. Additionally, we observe that the positions of the resonance fringes shift consistently with the change in $\Delta_1$, further supporting the interpretation that the biexciton is coherently populated through the two-pulse SUPER scheme.

The lower panel of Fig.~\ref{fig:Fig3}c displays the integrated PL intensities as a function of the second-pulse detuning $\Delta_2$. In agreement with the anisotropy measurements in Fig.~\ref{fig:Fig3}a, two distinct X resonance peaks are observed in the co-linearly polarized configuration (blue curve). Notably, only the resonance at $\Delta_2 = -6.1$ meV coincides with strong biexciton (XX) emission (red curve), confirming that this feature arises from XX-mediated excitation. As $\Delta_2$ is further decreased, the X/XX intensity ratio increases significantly, indicating that these excitation conditions lead to predominant neutral exciton preparation.
Although the spectral separation between the two resonances is small compared to the bandwidth of the excitation pulses, they can be reliably distinguished and followed over a broad range of $\Delta_{1,2}$ detunings, as shown in Fig.~\ref{fig:Fig3}d. Having confirmed coherent biexciton preparation via the SUPER protocol and clarified its impact on the properties of the exciton emission, we now investigate in more detail the changes in the observed SUPER resonances induced by rotating the laser polarization. 
In particular we focus on a peculiar excitation regime, where the SUPER protocol enables selective population of a single H or V-polarized exciton state, depending on the laser detuning and without altering its polarization.


\subsection{Emergence of an Orthogonal Exciton Resonance under Polarization-Misaligned SUPER}

\begin{figure*}
    \centering
    \includegraphics[width=0.85\textwidth]{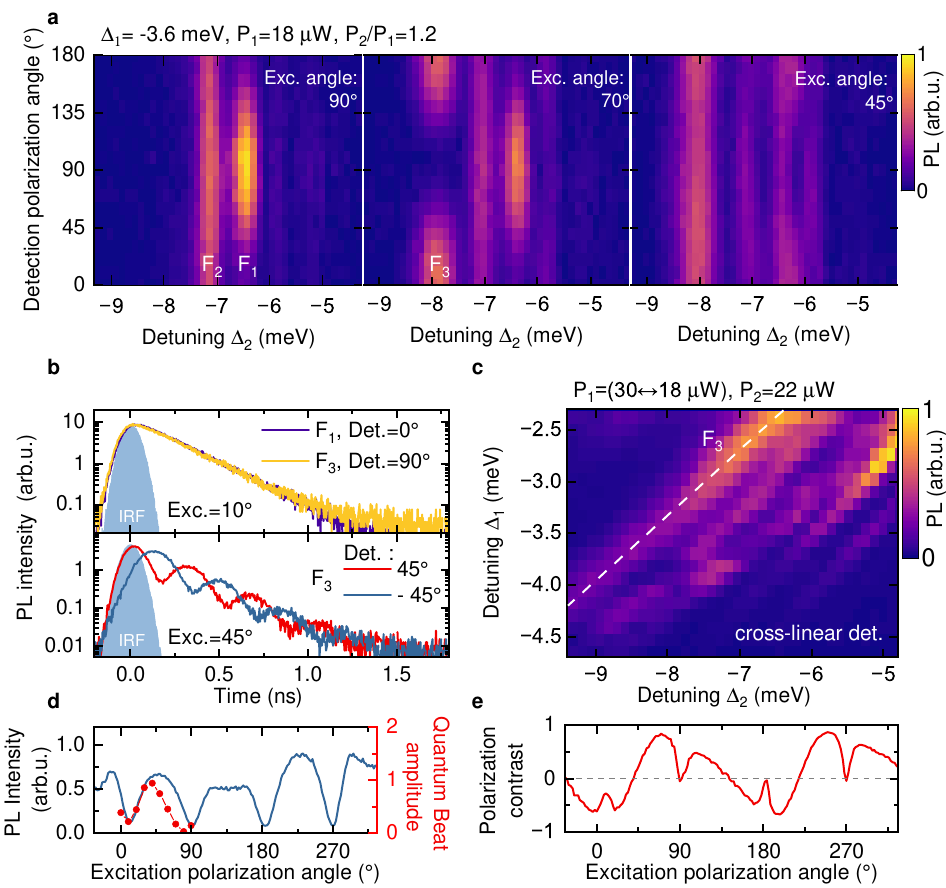}
    \caption{\textbf{a} Polarization anisotropy scan of the neutral exciton (X) photoluminescence (PL) under SUPER excitation as a function of $\Delta_2$ with the laser polarization angle aligned at 90$^\circ$, 70$^\circ$ and 45$^\circ$ with respect to the V exciton state. \textbf{b} (upper) X PL decay of the V/H-states for  $\Delta_2$ corresponding to either the F$_1$ or F$_3$ resonance fringe.   \textbf{b} (lower) Quantum beats observed in the X PL signal while exciting at the F$_3$ resonance fringe. \textbf{c} Intensity of the X PL under SUPER excitation while changing the detunings of the laser pulses for cross-linear polarization configuration. Dashed line indicate the F$_3$ resonance fringe related to efficient population of the orthogonal X state. (Experimental conditions and colorscale same as in Fig.~\ref{fig:Fig3}d). \textbf{d} Comparison of the X PL intensity under SUPER excitation for $\Delta_2$ corresponding to the F$_3$ resonance fringe and the observed quantum beat amplitude as a function of the polarization angle of the laser pulses. (QB amplitudes were measured for $\Delta_2$ tuned to the F$_1$ resonance fringe). \textbf{e} Linear polarization contrast ($\frac{I_{0^\circ}-I_{90^\circ }}{I_{0^\circ}+I_{90^\circ }}$) of X PL under SUPER for $\Delta_2$ corresponding to the F$_3$ excitation fringe.
    }

    \label{fig:Fig4}
\end{figure*}

In Fig.~\ref{fig:Fig4}a, we present polarization anisotropy measurements of the exciton (X) photoluminescence (PL) under SUPER excitation, shown as a function of $\Delta_2$ detuning for three selected excitation polarization angles. As expected, rotating the excitation polarization by 90$^\circ$ enables direct excitation of the orthogonal H-exciton state, similar to the behavior observed in Fig.~\ref{fig:Fig3}a, resulting in strongly polarized emission at $\Delta_2 \approx -6.5$~meV (F$_1$), and aligned with the excitation direction. Simultaneously, a second resonance fringe appears at $\Delta_2 \approx -7.2$~meV (F$_2$), corresponding to the indirect population of the neutral exciton via biexciton cascade, and yields unpolarized PL, in good agreement with the previous results shown in Fig.~\ref{fig:Fig3}a. Importantly, these excitation conditions remain relatively stable upon rotation of the excitation polarization.
Aligning the laser polarization between the H and V dipoles enables simultaneous excitation of both neutral exciton states. In this configuration, a coherent superposition of the H and V exciton states is driven, resulting in unpolarized PL emission. Moreover, since the interaction strength between the optical field and each dipole depends on their relative orientation, the resonant conditions for SUPER excitation are also modified. As a result, we observe a gradual decrease in both PL intensity and polarization degree at $\Delta_2 = -6.5$~meV as the excitation angle approaches 45$^\circ$. This orientation coincides with the strongest quantum beat amplitude in the time-resolved exciton PL (Suppl. Fig.~S15 and Fig.~\ref{fig:Fig4}d).

Interestingly, when the excitation polarization is not aligned with either the H- or V-exciton dipole, we observe the emergence of an additional resonance fringe at a larger detuning of $\Delta_2 \approx -8$~meV (F$_3$). As shown in the middle panel of Fig.~\ref{fig:Fig4}a, for an excitation angle of 70$^\circ$, the exciton PL corresponding to this detuning is strongly polarized, however, along the orthogonal, vertical direction ($\theta = 0^\circ$), whereas for $\Delta_2 \approx -6.5$~meV it remains polarized along the horizontal axis($\theta = 90^\circ$). This contrast diminishes for a 45$^\circ$ excitation angle, where both resonances result in unpolarized emission.

In the upper panel of Fig.~\ref{fig:Fig4}b, we observe that the V-polarized emission associated with the F$_3$ resonance fringe at $\Delta_2 \approx -8$~meV exhibits a similarly fast and monoexponential decay as that observed for direct SUPER excitation at $\Delta_2 = -6.9$~meV (Fig.~\ref{fig:Fig3}b). This behavior indicates that the neutral exciton emission does not result from a spontaneous biexciton cascade or relaxation from a higher-energy QD state. Furthermore, the lower panel of Fig.~\ref{fig:Fig4}b reveals pronounced quantum beat oscillations in the decay dynamics of the F$_3$ fringe when the excitation polarization is set to 45$^\circ$. This provides strong evidence that the resonance fringe corresponds to a coherent preparation of the neutral exciton - different from the biexciton-mediated excitation pathway, which, as shown in Fig.~\ref{fig:Fig3}b, does not 
produce quantum beats.

More evidence supporting the distinct nature of this excitation, likely addressing the mostly orthogonal exciton dipole, is presented in Fig.~\ref{fig:Fig4}c. The associated resonance fringe, highlighted by the white dashed line, can be consistently tracked across different combinations of $\Delta_{1,2}$ detunings. Notably, it exhibits a detuning dependence similar to that observed for the previously discussed SUPER-driven swing-up excitations of the X and XX states (cf. Fig.~\ref{fig:Fig3}d), further reinforcing its interpretation as a coherent and separate excitation pathway.

Although the working principle behind the intriguing third emission feature—namely, the emergence of exclusively V-polarized emission at larger detuning—is not fully understood and requires further theoretical investigation, several plausible scenarios can still be considered.

For one instance, this effect might be attributed to a slight misalignment of the excitation laser polarization with respect to the exciton dipole axes. For example, at an angle offset of approximately $\pm20^\circ$ from the H or V axis (see middle panel of Fig.~\ref{fig:Fig4}a), the excitation laser includes both H- and V-polarized components. In this configuration, the stronger H-polarized component could primarily excite the H-dipole, while the weaker V-polarized component might still contribute to the excitation of the V-dipole. However, this straightforward explanation falls short of accounting for the experimental observations. The V component, having a smaller projection onto the dipole moment, should result in substantially weaker PL compared to the H-polarized emission under similar conditions.
Contrary to this expectation, we observe that the intensity of the V-polarized emission at $\Delta_2 \approx -8$~meV is comparable to, or even rivals, that of the H-polarized emission at $\Delta_2 \approx -6.5$~meV. Moreover, a 70$^\circ$ orientation of the laser relates to a larger effective projection of the optical field on the H-dipole, however we detect no emission of this state at $\Delta_2\approx8$~meV. 
This strongly challenges the idea that the V-polarized emission arises merely from a residual V-component in a predominantly H-polarized excitation beam.
\newline
To further quantify this, we measure the total intensity of the emission associated with this resonance fringe as a function of the excitation polarization angle. These results are shown in Fig.~\ref{fig:Fig4}d, where, as a reference for the degree of dipole misalignment, we also plot the corresponding amplitude of the quantum beats extracted from the PL decay (Fig.~S16). Indeed, we observe that the related PL signal is strongly suppressed when the laser polarization is aligned with the H/V exciton dipoles, whereas it reaches largest values for excitation angles of $\phi=(2n+1)45^\circ$, similarly to the amplitude of the observed quantum beats. Notably, even a small $5^\circ$ misalignment can yield up to 50$\%$ of the maximal recorded intensity.
As evidenced, we show excitation conditions that enable selective coherent preparation of either H- or V-polarized exciton states without altering the polarization of the laser, simply by changing the detuning of the second pulse by approximately 2 meV. 
We verify the robustness of this effect by performing consecutive switching between the $\Delta_2$ detunings corresponding to the excitation of the respective H or V exciton states and recording no deviation up to 200 repetitions (Fig.~S17).

Remarkably, the coherent population of the mostly orthogonal exciton dipole persists over a broad range of excitation angles, as long as the laser polarization is not finely aligned along or directly in-between the H and V orientations. This is particularly well-visible in the Fig.~\ref{fig:Fig4}e, where we plot the polarization contrast of exciton emission corresponding to the related resonance fringe F$_\textrm{3}$ as a function of the orientation of the laser. 
We also emphasize that this phenomenon is not limited to the specific quantum dot or micropillar cavity investigated here. In principle, one could expect that the reported observations may be partly influenced by possible depolarization mechanisms arising from scattering of the laser on the microstructure. However, similar polarization properties of the three distinct resonance fringes associated with the SUPER excitation of the neutral exciton can be clearly observed for a quantum dot embedded in a planar structure from the same sample (Fig.~S18), demonstrating the robustness of this behavior.

Another potential explanation behind such an excitation is that it may be related to mechanisms involving the biexciton–exciton cascade. Under SUPER excitation of the XX state, even a small spectral overlap between the laser and the XX–X transition could potentially enhance the biexciton-to-exciton relaxation via a stimulated process. This would be analogous to stimulated two-photon excitation, resulting in a fast exciton decay and strongly polarized PL emission~\cite{Wei2022TailoringEmissions}. However, this does not explain the apparent preference for emission from the orthogonal exciton dipole—the one with a weaker coupling to the laser field.

Alternatively, the effect may be explained by a resonant XX-to-X relaxation driven by a SUPER-like process, resulting in a $\pi$-rotation of the Bloch vector within a dressed-state basis. Such a mechanism could lead to biexciton depopulation occurring within the duration of the laser pulse, when the energy levels are significantly modified due to AC-Stark shifts~\cite{calcagno2025dynamicallydressedstatesquantum}. In this picture, the biexciton PL line would be absent, as the XX state is coherently depleted before spontaneous emission can occur. This kind of behavior was, in particular, observed for the measurement presented in Fig.~\ref{fig:Fig3}c. There, we recorded a similar SUPER-driven resonance of the neutral exciton in a cross-polarized configuration (at $\Delta_2 = -6.9$~meV, not shown), yet no accompanying biexciton emission was detected. Furthermore, if the laser polarization is misaligned from the H and V dipole axes, the resulting asymmetry in dipole–laser coupling could shift the resonance conditions differently for each dipole, potentially explaining the observed selectivity with respect to $\Delta_2$.

These observations collectively point toward a more involved excitation mechanism than simple dipole projection or cascaded decay, suggesting a rich interplay between coherent dynamics, laser detuning, and polarization selection in the SUPER excitation scheme. While our experimental findings clearly demonstrate the robust and tunable coherent preparation of both exciton dipoles—including selective excitation of the orthogonal dipole without modifying laser polarization—the underlying physical mechanism remains to be fully understood.

Further theoretical modeling, particularly incorporating dressed-state dynamics and ultrafast population transfer pathways under AC-Stark conditions, will be essential to provide a complete microscopic picture of this process. Nonetheless, our results establish a novel degree of control over neutral exciton states in quantum dots, enabled by the SUPER protocol, and open new avenues for deterministic state preparation in solid-state quantum photonic devices.
Lastly, we note that the selective nature of the SUPER protocol also enables excitation of other excitonic complexes, such as trions, by appropriately tuning the laser detuning. While a detailed investigation is beyond the scope of this work, our experiments reveal clear trion emission under alternative excitation conditions (see Supplementary Fig. S19). These observations are consistent with recent experimental demonstration of coherent trion preparation via SUPER~\cite{Nawrath2019a, Boos2024CoherentSwingUp}, and point toward exciting prospects for scalable multi-photon protocols and deterministic cluster state generation~\cite{LindlerPRL09}.


\section{Conclusion}

In summary, we have conducted a comprehensive experimental study of swing-up resonant excitation (SUPER) in a GaAs quantum dot, demonstrating efficient and coherent control of exciton and, for the first time, biexciton states. By tuning pulse parameters and polarization, we identify distinct SUPER resonances that enable selective excitation of H or V exciton dipoles without changing the laser polarization.
These results establish SUPER as a powerful and versatile scheme for state-selective excitation, offering a promising path toward high-fidelity entangled photon pair generation via the XX–X cascade. The method’s off-resonant nature and polarization flexibility make it well-suited for scalable quantum photonic applications.


\section{Methodology}

\textbf{Optical characterization}
To perform photoluminescence measurements, the sample was placed in a closed-cycle helium cryostat (attoDRY 800, attocube systems AG) with a base temperature of 4 K. The cryostat was equipped with a microscope objective to focus the laser light onto the sample (NA=0.81, 60×, attocube systems AG). For above bandgap excitation measurements a 650 nm laser was utilized either in CW or pulsed (80 MHz) mode (Picoquant LDH series). To enable various excitation schemes, we employed a tunable OPO module (Chameleon Compact OPO-Vis, Coherent Inc.) pumped by a Ti:Sapphire laser (Chameleon Ultra II, Coherent Inc.) producing 150 fs pulses at an 80 MHz repetition rate. To further optimize the laser spectral window for various excitation methods, we also employed a home-built 4f pulse shaper (f=500 mm) equipped with an 1800 groove/mm diffraction grating and a programmable single-mask transmissive spatial light modulator (SLM-S320, Jenoptik) in the Fourier plane. The SLM consisting of 320 pixels, helps to carve a narrow spectral range with a resolution of FWHM$\approx$0.08 nm corresponding to a single px transmission. For acquiring  emission spectra, we utilized a fiber-coupled spectrometer (Horiba iHR 550) equipped with a 1200 groove/mm grating and a CCD camera (Syncerity Bi UV-Vis Horiba). For second-order correlation measurements, we employed a fiber-based Hanbury Brown and Twiss (HBT) setup coupled to superconducting nanowire single-photon detectors (SNSPD, ID218, ID Quantique) and time tagger (Swabian Instruments). Further details on the experimental setup are provided in Fig.~S2.

\textbf{Autocorrelation measurements}
For the presented results of the autocorrelation measurements of the exciton PL under different excitation schemes the extracted $g^{(2)}(0)$ values were calculated as the area of the zero-delay peak divided by the average of the areas of the peaks at the largest measured delay times. The error range reported was the standard deviation calculated by error propagation, assuming a Poisson statistic for the peaks' areas. More details available in SI note.  

\textbf{Theory}. To model the state preparation dynamics and the PL intensity, we consider a four-level model that includes the ground state $\ket{G}$, two orthogonal exciton states $\ket{X_V}$ and $\ket{X_H}$ with linear polarization, and the biexciton state $\ket{XX}$. Their respective energies are $E_G = 0$, $E_{X_V} = \hbar \qty(\omega_0 - \frac 1 2 \Delta_{\rm FSS})$, $E_{X_H} = \hbar \qty(\omega_0 + \frac 1 2 \Delta_{\rm FSS})$, and $E_{XX} = \qty(2 \hbar \omega_0 - E_b)$, where $\hbar \Delta_{\rm FSS}$ is the fine structure splitting (FSS), and $E_b$ is the biexciton binding energy. 
Interaction with the laser pulse is modelled with the conventional dipole and rotating wave approximations. 
Adopting an interaction picture with respect to $H_0 = \qty(2 \hbar \omega_0 - E_b) \dyad{XX} + \hbar \omega_0 \qty(\dyad{X_H} + \dyad{X_V})$, the system Hamiltonian is 
\begin{multline}
\label{eq:Hamiltonian}
    H(t) = \frac{\hbar \Delta_{\rm FSS}}{2} \qty(\dyad{X_H} - \dyad{X_V}) \\
    + \sum_{j=1,2} \frac{\hbar}{2} \Omega_j(t) \qty[ e^{-i \delta_j t} \qty(\dyad{X_H}{G} + e^{-i \frac{E_b}{\hbar} t} \dyad{XX}{X_H}) + \mathrm{h.c.}] ,
\end{multline}
where $\Omega_j(t)$ is the temporal envelope of the $j$-th pulse and $\delta_j$ its frequency detuning with respect to $\omega_0$. The corresponding energy detuning is $\Delta_j = \hbar \delta_j$.
In Eq.~\eqref{eq:Hamiltonian}, we assume that the laser polarization is aligned with the H exciton. An arbitrary excitation angle can be modeled by rotating the exciton state from $\ket{X_H}$ to $\ket{X_H'} = \cos(\phi) \ket{X_H} - \sin(\phi) \ket{X_V}$.  

The dynamics of the density operator is obtained by solving the master equation
\begin{equation}
    \label{eq:master_equation}
    \dv{t} \rho = -\frac{i}{\hbar} \comm{H(t)}{\rho} + \Gamma \sum_{p=H, V} \qty(\mathcal{L}_{\dyad{G}{X_p}}[\rho] + \mathcal{L}_{\dyad{X_p}{B}} [\rho]) ,
\end{equation}
which includes spontaneous emission via a set of Lindblad super-operators $\mathcal{L}_{A}[\rho] = A \rho A^{\dag} - \frac{1}{2} \acomm{A^\dag A}{\rho}$. Here, we consider identical emission rate $\Gamma = (200 \, \mathrm{ps})^{-1}$ for all transitions.
Equation \eqref{eq:master_equation} is solved with a 4th-order Runge-Kutta method with initial condition $\rho(t_0) = \dyad{G}$.
Then, the number of H-polarized photons emitted in the transition from $\ket{X_H} \to \ket{G}$ is calculated as
\begin{equation}
    N_H = \Gamma \int_{t_0}^{+\infty} \dd{t} \Tr \qty[ \dyad{X_H} \rho(t)] .
\end{equation}
The number of photons emitted in the opposite polarization is $N_V = \Gamma \int_{t_0}^{+\infty} \dd{t} \Tr \qty[ \dyad{X_V} \rho(t)]$.

For the temporal shape of the pulses, we replace the conventional Gaussian envelope (see e.g.\ Ref.~\cite{Vannucci2023HighlyExcitation}) with
\begin{equation}
    \Omega_j(t) = \frac{\Theta_j}{t_p \pi \mathrm{Erf}(r/2)} \mathrm{Sinc}\qty(\frac{t}{t_p \pi}) \mathrm{Exp} \qty[-\qty(\frac{t}{r t_p})^2],
\end{equation}
where $\Theta_j$ is the pulse area, and $r$ is a free numerical parameter that we fix to $r = 10$.
The Fourier-transformed pulse amplitude in the frequency domain is
\begin{align}
    \widehat \Omega_j(\omega) & = \frac{\Theta_j}{2 \mathrm{Erf}(r/2)} \mathrm{Erf} \qty[\frac r 2 \qty(\omega t_p - \delta_j t_p + 1)] \nonumber \\
    & \quad - \frac{\Theta_j}{2 \mathrm{Erf}(r/2)} \mathrm{Erf} \qty[\frac r 2 \qty(\omega t_p - \delta_j t_p - 1)] ,
\end{align}
corresponding to square pulses centred at $\delta_j$ with smooth corners. The corresponding pulse power is $|\widehat \Omega_j(\omega)|^2$, and the FWHM in wavelength for such a pulse is ${\rm FWHM}_\lambda = 0.923 \lambda_0^2 / (\pi c t_p)$, with $\lambda_0$ the central wavelength.

The incoming pulse power $P^{(\rm in)}_j$ is measured above the sample and depends on the amplitude of the incoming electric field $E^{(\rm in)}_j$ and the pulse spectral width $W_j$ as $P^{(\rm in)}_j \propto |E^{(\rm in)}_j|^2 W_j$.
In contrast, the effective pulse area $\Theta_j$ used in the simulation satisfies $\Theta_j^2 \propto |E^{(\rm QD)}_j|^2 W_j$, with $E^{(\rm QD)}_j$ the electric field amplitude at the QD position.
The latter depends on the transmission through the top DBR mirror, which varies as a function of the detuning $\Delta_j$.
To take this into account, we use the same numerical method as in Ref.~\cite{Madigawa2025DeterministicSources} to calculate the ratio $F_j = |E^{(\rm QD)}_j(\Delta_j) / E^{(\rm in)}_j (\Delta_j)|$ between the field at the QD position and the incoming field. 
Then, for a given incoming power ratio we have
\begin{equation}
    \frac{P^{(\rm in)}_2}{P^{(\rm in)}_1} = \frac{|E^{(\rm in)}_2|^2 W_2}{|E^{(\rm in)}_1|^2 W_1} = \qty(\frac{\Theta_2}{\Theta_1} \frac{F_1}{F_2})^2 \frac{W_2}{W_1},
\end{equation}
from which we extract the relative pulse areas $\Theta_j$ to be used in the simulation. 


\section*{Acknowledgements}
The authors acknowledge the cleanroom facilities at DTU Nanolab – National Centre for Nano Fabrication and Characterization. 

\section*{Funding}

The authors acknowledge support from the European Research Council (ERC-CoG ``Unity", grant no. 865230 and ERC-StG ``TuneTMD", grant no. 101076437), the Villum Foundation (grant no. VIL53033), the Innovation Fund Denmark (QLIGHT, no. 4356-00002A), the TICRA foundation, and the Carlsberg Foundation (grant no. CF21-0496). LV acknowledges support from the Novo Nordisk Foundation (grant no. NNF24OC0094739 ``SUPER-Q''). Ar.R. acknowledges support of the EU HE EIC Pathfinder challenges action under grant agreement No. 101115575 (Q-ONE), from the QuantERA program via the project MEEDGARD (FFG Grant No. 906046) the Austrian Science Fund FWF via the Research Group FG5 [10.55776/FG5] and the cluster of excellence quantA [10.55776/COE1] as well as the Linz Institute of Technology (LIT), and the LIT Secure and Correct Systems Lab, supported by the State of Upper Austria. VR and GW acknowledge funding from FWF via DarkEneT [10.55776/TAI556] and Cluster of Excellence Quantum Science Austria [10.55776/COE1]. 

\section*{Author Contributions}
C.P. and Al.R. performed the optical characterization and the data analysis, and processing of all the data. A.G., S.C.S., and Ar.R. grew the QD samples. A.M. fabricated the samples. L.V. modeled the quantum dynamics and calculated the state preparation fidelity and the theoretical PL intensity. MAJ performed optical simulation of the micropillar device and calculated the field intensity at the position of the emitter. C.P. and V.R. assembled a home-built 4f pulse-shaping setup using the SLM. G.W. provided fruitful discussion and valuable suggestions on the manuscript. 
N.G., Ar.R., and B.M. conceived the idea and coordinated the project. C.P., Al.R., V.R., L.V., Ar.R., N.G. and B.M. wrote the manuscript with the input of all co-authors.

\section*{Competing Interests}
The authors declare no competing financial interest.

\section*{Data and materials availability}
\noindent
All data needed to verify the results and conclusions of this publication are present in the main paper and the Supplementary Materials.

\section*{References}

\bibliography{bibliography}

@article{Madigawa2025DeterministicSources,
author = {Madigawa, Abdulmalik A. and Jacobsen, Martin Arentoft and Piccinini, Claudia and Wyborski, Paweł and Garcia Jr., Ailton and da Silva, Saimon F. Covre and Rastelli, Armando and Munkhbat, Battulga and Gregersen, Niels},
title = {Deterministic Fabrication of GaAs-Quantum-Dot Micropillar Single-Photon Sources},
journal = {Advanced Quantum Technologies},
volume = {8},
number = {8},
pages = {e2500128},
doi = {https://doi.org/10.1002/qute.202500128},
url = {https://advanced.onlinelibrary.wiley.com/doi/abs/10.1002/qute.202500128},
year = {2025}
}

@article{Vannucci2023HighlyExcitation,
  title = {Highly efficient and indistinguishable single-photon sources via phonon-decoupled two-color excitation},
  author = {Vannucci, Luca and Gregersen, Niels},
  journal = {Phys. Rev. B},
  volume = {107},
  issue = {19},
  pages = {195306},
  numpages = {11},
  year = {2023},
  month = {May},
  publisher = {American Physical Society},
  doi = {10.1103/PhysRevB.107.195306},
  url = {https://link.aps.org/doi/10.1103/PhysRevB.107.195306}
}

@article{Bracht2023Dressed-stateSchemes,
  title = {Dressed-state analysis of two-color excitation schemes},
  author = {Bracht, Thomas K. and Seidelmann, Tim and Karli, Yusuf and Kappe, Florian and Remesh, Vikas and Weihs, Gregor and Axt, Vollrath Martin and Reiter, Doris E.},
  journal = {Phys. Rev. B},
  volume = {107},
  issue = {3},
  pages = {035425},
  numpages = {11},
  year = {2023},
  month = {Jan},
  publisher = {American Physical Society},
  doi = {10.1103/PhysRevB.107.035425},
  url = {https://link.aps.org/doi/10.1103/PhysRevB.107.035425}
}

@article{Boos2024CoherentSwingUp,
author = {Boos, Katarina and Sbresny, Friedrich and Kim, Sang Kyu and Kremser, Malte and Riedl, Hubert and Bopp, Frederik W. and Rauhaus, William and Scaparra, Bianca and Jöns, Klaus D. and Finley, Jonathan J. and Müller, Kai and Hanschke, Lukas},
title = {Coherent Swing-Up Excitation for Semiconductor Quantum Dots},
journal = {Advanced Quantum Technologies},
volume = {7},
number = {4},
pages = {2300359},
doi = {https://doi.org/10.1002/qute.202300359},
url = {https://advanced.onlinelibrary.wiley.com/doi/abs/10.1002/qute.202300359},
year = {2024}
}

@article{Karli2022SUPEREmitter,
author = {Karli, Yusuf and Kappe, Florian and Remesh, Vikas and Bracht, Thomas K. and M{\"u}nzberg, Julian and Covre da Silva, Saimon and Seidelmann, Tim and Axt, Vollrath Martin and Rastelli, Armando and Reiter, Doris E. and Weihs, Gregor},
title = {SUPER Scheme in Action: Experimental Demonstration of Red-Detuned Excitation of a Quantum Emitter},
journal = {Nano Letters},
volume = {22},
number = {16},
pages = {6567-6572},
year = {2022},
doi = {10.1021/acs.nanolett.2c01783},
URL = {https://doi.org/10.1021/acs.nanolett.2c01783},
}

@article{Nawrath2019a,
    author = {Nawrath, C. and Olbrich, F. and Paul, M. and Portalupi, S. L. and Jetter, M. and Michler, P.},
    title = {Coherence and indistinguishability of highly pure single photons from non-resonantly and resonantly excited telecom C-band quantum dots},
    journal = {Applied Physics Letters},
    volume = {115},
    number = {2},
    pages = {023103},
    year = {2019},
    month = {07},
    issn = {0003-6951},
    doi = {10.1063/1.5095196},
    url = {https://doi.org/10.1063/1.5095196},

    
}

@article{Somaschi2016,
    title = {{Near-optimal single-photon sources in the solid state}},
    year = {2016},
    journal = {Nature Photonics},
    author = {Somaschi, N. and Giesz, V. and De Santis, L. and Loredo, J. C. and Almeida, M. P. and Hornecker, G. and Portalupi, S. L. and Grange, T. and Ant{\'{o}}n, C. and Demory, J. and G{\'{o}}mez, C. and Sagnes, I. and Lanzillotti-Kimura, N. D. and Lema{\'{i}}tre, A. and Auffeves, A. and White, A. G. and Lanco, L. and Senellart, P.},
    number = {5},
    pages = {340--345},
    volume = {10},
    doi = {10.1038/nphoton.2016.23},
    issn = {17494893},
    arxivId = {1510.06499}
}

@article{Ding2016,
    title = {{On-Demand Single Photons with High Extraction Efficiency and Near-Unity Indistinguishability from a Resonantly Driven Quantum Dot in a Micropillar}},
    year = {2016},
    journal = {Physical Review Letters},
    author = {Ding, Xing and He, Yu and Duan, Z. C. and Gregersen, Niels and Chen, M. C. and Unsleber, S. and Maier, S. and Schneider, Christian and Kamp, Martin and H{\"{o}}fling, Sven and Lu, Chao Yang and Pan, Jian Wei},
    number = {2},
    pages = {1--6},
    volume = {116},
    doi = {10.1103/PhysRevLett.116.020401},
    issn = {10797114},
    arxivId = {1601.00284}
}

@article{Bracht2021a,
    title = {{Swing-Up of Quantum Emitter Population Using Detuned Pulses}},
    year = {2021},
    journal = {PRX Quantum},
    author = {Bracht, Thomas K. and Cosacchi, Michael and Seidelmann, Tim and Cygorek, Moritz and Vagov, Alexei and Axt, V. Martin and Heindel, Tobias and Reiter, Doris E.},
    number = {4},
    pages = {1},
    volume = {2},
    publisher = {American Physical Society},
    url = {https://doi.org/10.1103/PRXQuantum.2.040354},
    doi = {10.1103/PRXQuantum.2.040354},
    issn = {26913399},
    arxivId = {2111.10236}
}

@article{Wang2019,
    title = {{Towards optimal single-photon sources from polarized microcavities}},
    year = {2019},
    journal = {Nature Photonics},
    author = {Wang, Hui and He, Yu Ming and Chung, T. H. and Hu, Hai and Yu, Ying and Chen, Si and Ding, Xing and Chen, M. C. and Qin, Jian and Yang, Xiaoxia and Liu, Run Ze and Duan, Z. C. and Li, J. P. and Gerhardt, S. and Winkler, K. and Jurkat, J. and Wang, Lin Jun and Gregersen, Niels and Huo, Yong Heng and Dai, Qing and Yu, Siyuan and H{\"{o}}fling, Sven and Lu, Chao Yang and Pan, Jian Wei},
    number = {11},
    pages = {770--775},
    volume = {13},
    isbn = {4156601904943},
    doi = {10.1038/s41566-019-0494-3},
    issn = {17494893}
}

@article{Madigawa2023AssessingDevices,
    title = {{Assessing the alignment accuracy of state-of-the-art deterministic fabrication methods for single quantum dot devices}},
    year = {2023},
    journal = {ACS Photonics},
    author = {Madigawa, Abdulmalik A. and Donges, Jan N. and Ga{\'{a}}l, Benedek and Li, Shulun and Liu, Hanqing and Dai, Deyan and Su, Xiangbin and Shang, Xiangjun and Ni, Haiqiao and Schall, Johannes and Rodt, Sven and Niu, Zhichuan and Gregersen, Niels and Reitzenstein, Stephan and Munkhbat, Battulga},
    month = {9},
    pages = {1012--1023},
    volume = {11},
    publisher = {American Chemical Society},
    url = {http://arxiv.org/abs/2309.14795},
    doi = {10.1021/ACSPHOTONICS.3C01368/ASSET/IMAGES/LARGE/PH3C01368{\_}0004.JPEG},
    issn = {23304022}
}

@article{Thomas2021BrightDipole,
    title = {{Bright Polarized Single-Photon Source Based on a Linear Dipole}},
    year = {2021},
    journal = {Physical Review Letters},
    author = {Thomas, S. E. and Billard, M. and Coste, N. and Wein, S. C. and {Priya} and Ollivier, H. and Krebs, O. and Taza{\"{i}}rt, L. and Harouri, A. and Lemaitre, A. and Sagnes, I. and Anton, C. and Lanco, L. and Somaschi, N. and Loredo, J. C. and Senellart, P.},
    number = {23},
    month = {6},
    pages = {233601},
    volume = {126},
    publisher = {American Physical Society},
    url = {https://journals.aps.org/prl/abstract/10.1103/PhysRevLett.126.233601},
    doi = {10.1103/PHYSREVLETT.126.233601/FIGURES/4/MEDIUM},
    issn = {10797114},
    pmid = {34170172},
    arxivId = {2007.04330}
}

@article{Reindl2019HighlyDots,
    title = {{Highly indistinguishable single photons from incoherently excited quantum dots}},
    year = {2019},
    journal = {Physical Review B},
    author = {Reindl, Marcus and Weber, Jonas H. and Huber, Daniel and Schimpf, Christian and Covre Da Silva, Saimon F. and Portalupi, Simone L. and Trotta, Rinaldo and Michler, Peter and Rastelli, Armando},
    number = {15},
    month = {10},
    pages = {155420},
    volume = {100},
    publisher = {American Physical Society},
    url = {https://journals.aps.org/prb/abstract/10.1103/PhysRevB.100.155420},
    doi = {10.1103/PHYSREVB.100.155420/FIGURES/4/MEDIUM},
    issn = {24699969}
}

@article{Uppu2020On-chipSource,
    title = {{On-chip deterministic operation of quantum dots in dual-mode waveguides for a plug-and-play single-photon source}},
    year = {2020},
    journal = {Nature Communications},
    author = {Uppu, Ravitej and Eriksen, Hans T. and Thyrrestrup, Henri and U{\u{g}}urlu, Aslı D. and Wang, Ying and Scholz, Sven and Wieck, Andreas D. and Ludwig, Arne and L{\"{o}}bl, Matthias C. and Warburton, Richard J. and Lodahl, Peter and Midolo, Leonardo},
    number = {1},
    month = {7},
    pages = {1--6},
    volume = {11},
    publisher = {Nature Publishing Group},
    url = {https://www.nature.com/articles/s41467-020-17603-9},
    doi = {10.1038/s41467-020-17603-9},
    issn = {2041-1723},
    pmid = {32728025},
}

@article{Schweickert2018On-demandSource,
    title = {{On-demand generation of background-free single photons from a solid-state source}},
    year = {2018},
    journal = {Applied Physics Letters},
    author = {Schweickert, Lucas and J{\"{o}}ns, Klaus D. and Zeuner, Katharina D. and Covre Da Silva, Saimon Filipe and Huang, Huiying and Lettner, Thomas and Reindl, Marcus and Zichi, Julien and Trotta, Rinaldo and Rastelli, Armando and Zwiller, Val},
    number = {9},
    month = {2},
    pages = {93106},
    volume = {112},
    publisher = {American Institute of Physics Inc.},
    url = {/aip/apl/article/112/9/093106/36016/On-demand-generation-of-background-free-single},
    doi = {10.1063/1.5020038/13034201/093106{\_}1{\_}ONLINE.PDF},
    issn = {00036951},
    arxivId = {1712.06937}
}

@article{Muller2014On-demandPairs,
    title = {{On-demand generation of indistinguishable polarization-entangled photon pairs}},
    year = {2014},
    journal = {Nature Photonics},
    author = {M{\"{u}}ller, M. and Bounouar, S. and J{\"{o}}ns, K. D. and Gl{\"{a}}ssl, M. and Michler, P.},
    number = {3},
    month = {2},
    pages = {224--228},
    volume = {8},
    publisher = {Nature Publishing Group},
    url = {https://www.nature.com/articles/nphoton.2013.377},
    doi = {10.1038/nphoton.2013.377},
    issn = {1749-4893},
    arxivId = {1308.4257}
}

@article{Ates2009Post-SelectedMicrocavity,
    title = {{Post-Selected Indistinguishable Photons from the Resonance Fluorescence of a Single Quantum Dot in a Microcavity}},
    year = {2009},
    journal = {Physical Review Letters},
    author = {Ates, S. and Ulrich, S. M. and Reitzenstein, S. and L{\"{o}}ffler, A. and Forchel, A. and Michler, P.},
    number = {16},
    month = {10},
    pages = {167402},
    volume = {103},
    publisher = {American Physical Society},
    url = {https://journals.aps.org/prl/abstract/10.1103/PhysRevLett.103.167402},
    doi = {10.1103/PHYSREVLETT.103.167402/FIGURES/4/MEDIUM},
    issn = {00319007}
}

@article{Heindel2023QuantumTechnology,
    title = {{Quantum dots for photonic quantum information technology}},
    year = {2023},
    journal = {Advances in Optics and Photonics},
    author = {Heindel, Tobias and Kim, Je-Hyung and Gregersen, Niels and Rastelli, Armando and Reitzenstein, Stephan},
    number = {3},
    month = {9},
    pages = {613},
    volume = {15},
    url = {https://opg.optica.org/abstract.cfm?URI=aop-15-3-613},
    doi = {10.1364/AOP.490091},
    issn = {1943-8206}
}

@article{Ollivier2020ReproducibilitySources,
    title = {{Reproducibility of High-Performance Quantum Dot Single-Photon Sources}},
    year = {2020},
    journal = {ACS Photonics},
    author = {Ollivier, Hélène and Maillette De Buy Wenniger, Ilse and Thomas, Sarah and Wein, Stephen C. and Harouri, Abdelmounaim and Coppola, Guillaume and Hilaire, Paul and Millet, Clément and Lema{\^{i}}tre, Aristide and Sagnes, Isabelle and Krebs, Olivier and Lanco, Loïc and Loredo, Juan C. and Ant{\'{o}}n, Carlos and Somaschi, Niccolo and Senellart, Pascale},
    number = {4},
    month = {4},
    pages = {1050--1059},
    volume = {7},
    publisher = {American Chemical Society},
    url = {https://pubs.acs.org/doi/full/10.1021/acsphotonics.9b01805},
    doi = {10.1021/ACSPHOTONICS.9B01805/ASSET/IMAGES/LARGE/PH9B01805{\_}0004.JPEG},
    issn = {23304022},
    arxivId = {1910.08863}
}

@article{Wei2022TailoringEmissions,
    title = {{Tailoring solid-state single-photon sources with stimulated emissions}},
    year = {2022},
    journal = {Nature Nanotechnology 2022 17:5},
    author = {Wei, Yuming and Liu, Shunfa and Li, Xueshi and Yu, Ying and Su, Xiangbin and Li, Shulun and Shang, Xiangjun and Liu, Hanqing and Hao, Huiming and Ni, Haiqiao and Yu, Siyuan and Niu, Zhichuan and Iles-Smith, Jake and Liu, Jin and Wang, Xuehua},
    number = {5},
    month = {4},
    pages = {470--476},
    volume = {17},
    publisher = {Nature Publishing Group},
    url = {https://www.nature.com/articles/s41565-022-01092-6},
    doi = {10.1038/s41565-022-01092-6},
    issn = {1748-3395},
    pmid = {35410369},
    arxivId = {2109.09284}
}

@misc{calcagno2025dynamicallydressedstatesquantum,
      title={Dynamically Dressed States of a Quantum Four-Level System}, 
      author={Carolin Calcagno and Friedrich Sbresny and Thomas K. Bracht and Sang Kyu Kim and Eduardo Zubizarreta Casalengua and Katarina Boos and William Rauhaus and Hubert Riedl and Jonathan J. Finley and Doris E. Reiter and Kai Müller},
      year={2025},
      eprint={2506.14643},
      archivePrefix={arXiv},
      primaryClass={quant-ph},
      url={https://arxiv.org/abs/2506.14643}, 
}

@article{KoongPRL21,
  title = {Coherent Dynamics in Quantum Emitters under Dichromatic Excitation},
  author = {Koong, Z. X. and Scerri, E. and Rambach, M. and Cygorek, M. and Brotons-Gisbert, M. and Picard, R. and Ma, Y. and Park, S. I. and Song, J. D. and Gauger, E. M. and Gerardot, B. D.},
  journal = {Phys. Rev. Lett.},
  volume = {126},
  issue = {4},
  pages = {047403},
  numpages = {7},
  year = {2021},
  month = {Jan},
  publisher = {American Physical Society},
  doi = {10.1103/PhysRevLett.126.047403},
  url = {https://link.aps.org/doi/10.1103/PhysRevLett.126.047403}
}

@article{HeNatPhys19,
	title = {Coherently driving a single quantum two-level system with dichromatic laser pulses},
	volume = {15},
	copyright = {2019 The Author(s), under exclusive licence to Springer Nature Limited},
	issn = {1745-2481},
	url = {https://www.nature.com/articles/s41567-019-0585-6},
	doi = {10.1038/s41567-019-0585-6},
	number = {9},
	urldate = {2025-10-07},
	journal = {Nature Physics},
	author = {He, Yu-Ming and Wang, Hui and Wang, Can and Chen, M.-C. and Ding, Xing and Qin, Jian and Duan, Z.-C. and Chen, Si and Li, J.-P. and Liu, Run-Ze and Schneider, C. and Atatüre, Mete and Höfling, Sven and Lu, Chao-Yang and Pan, Jian-Wei},
	month = sep,
	year = {2019},
	pages = {941--946},
	}

@misc{yan2025robustentangledphotongeneration,
      title={Robust entangled photon generation enabled by single-shot Floquet driving}, 
      author={Jun-Yong Yan and Paul C. A. Hagen and Hans-Georg Babin and Wei E. I. Sha and Andreas D. Wieck and Arne Ludwig and Chao-Yuan Jin and Vollrath M. Axt and Da-Wei Wang and Moritz Cygorek and Feng Liu},
      year={2025},
      eprint={2504.02753},
      archivePrefix={arXiv},
      primaryClass={quant-ph},
      url={https://arxiv.org/abs/2504.02753}, 
}

@article{FlissikowskiPRL01,
  title = {Photon Beats from a Single Semiconductor Quantum Dot},
  author = {Flissikowski, T. and Hundt, A. and Lowisch, M. and Rabe, M. and Henneberger, F.},
  journal = {Phys. Rev. Lett.},
  volume = {86},
  issue = {14},
  pages = {3172--3175},
  numpages = {0},
  year = {2001},
  month = {Apr},
  publisher = {American Physical Society},
  doi = {10.1103/PhysRevLett.86.3172},
  url = {https://link.aps.org/doi/10.1103/PhysRevLett.86.3172}
}

@article{LindlerPRL09,
  title = {Proposal for Pulsed On-Demand Sources of Photonic Cluster State Strings},
  author = {Lindner, Netanel H. and Rudolph, Terry},
  journal = {Phys. Rev. Lett.},
  volume = {103},
  issue = {11},
  pages = {113602},
  numpages = {4},
  year = {2009},
  month = {Sep},
  publisher = {American Physical Society},
  doi = {10.1103/PhysRevLett.103.113602},
  url = {https://link.aps.org/doi/10.1103/PhysRevLett.103.113602}
}

@article{WilburAPLphot22,
    author = {Wilbur, G. R. and Binai-Motlagh, A. and Clarke, A. and Ramachandran, A. and Milson, N. and Healey, J. P. and O’Neal, S. and Deppe, D. G. and Hall, K. C.},
    title = {Notch-filtered adiabatic rapid passage for optically driven quantum light sources},
    journal = {APL Photonics},
    volume = {7},
    number = {11},
    pages = {111302},
    year = {2022},
    month = {11},
    issn = {2378-0967},
    doi = {10.1063/5.0090048},
    url = {https://doi.org/10.1063/5.0090048},
}

\end{document}